\documentclass[aps,twocolumn,showpacs,prl]{revtex4}
\usepackage{amssymb}
\usepackage{natbib}
\usepackage{amsmath}
\usepackage{amsfonts}
\usepackage{graphicx}
\usepackage{mathrsfs}
\usepackage{dcolumn}
\usepackage{bm}
\usepackage{color}
\usepackage{cancel}
\usepackage{mathbbol}
\usepackage{epsfig}
\usepackage{units}

\definecolor{rot}{rgb}{0.75,0.05,0.25}
\definecolor{hellgrau}{gray}{0.5}
\definecolor{blau}{rgb}{0,0,0.7}

\begin{document}

\title{Reply to W. G. Hoover [arXiv:1204.0312v2]}
\author{Michele Campisi, Fei Zhan, Peter Talkner, and Peter H\"anggi}
\affiliation{Institute of Physics, University of Augsburg,
  Universit\"atsstr. 1, D-86135 Augsburg, Germany}
\date{\today }

\begin{abstract}
In response to W. G. Hoover's comment [arXiv:1204.0312v2] on our work [arXiv:1203.5968], we
show explicitly that the divergence of the velocity field associated with the Nos\'e-Hoover equations is nonzero,
implying that those equations are  {\it not} volume preserving, and hence, as often stated in the literature, are not Hamiltonian.
We further elucidate that  the trajectories $\{q(t)\}$ generated by the
Nos\'e-Hoover equations are generally \emph{not identical} to those generated by Dettmann's Hamiltonian.
Dettmann's Hamiltonian produces the same trajectories as the Nos\'e-Hoover equations \emph{only on a  specific energy shell},
but not on the neighboring ones. This fact explains why the Nos\'e-Hoover equations are not volume preserving.
The Hamiltonian that we put forward with [arXiv:1203.5968] instead produces thermostated dynamics irrespective of the energy value. The main advantage of our Hamiltonian thermostat over previous ones is that it contains kinetic energy terms that are of standard form with coordinate-independent masses and
consequently is readily matched in laboratory experiments.

\end{abstract}


 \maketitle

In his comment \cite{Hoover} to our preprint \cite{Campisi} W. G. Hoover criticizes our statement
that \emph{the Nos\'e-Hoover equations are not Hamiltonian}, and maintains that, on the contrary, they are Hamiltonian.
We found the arguments given by  Hoover in support of his statement incorrect.

Hoover \cite{Hoover} begins with Nos\'e's Hamiltonian of a harmonic oscillator of unit mass and unit frequency:
\begin{equation}
H_{\rm Nos\acute{e}} = (1/2)[  (p/s)^2 + q^2 + (p_s/\tau)^2 ] + T\ln s
\label{Hoover1}
 \end{equation}
where $\tau$ and $T$ are parameters controlling a time scale and the temperature of the thermostat, respectively.
The corresponding equations of motion:
\begin{equation}
\dot q = p/s^2 \ ; \ \dot p = -q \ ; \ \dot s = p_s/\tau^2 \ ; \ \dot p_s = p^2/s^3 - T/s \ ,
\end{equation}
are by definition Hamiltonian \cite{Nose}. Hoover then proceeds with a ``time-rescaling'' to obtain the new set:
\begin{equation}
\dot q = p/s \ ; \ \dot p = -sq \ ; \ \dot s = s p_s/\tau^2 \ ; \ \dot p_s = (p/s)^2 - T.
\label{Hoover3}
\end{equation}
This new set is evidently non-Hamiltonian. A crucial
property of a Hamiltonian flow is that the divergence of the
velocity field in phase space is null (this is equivalent to saying that
the flow is incompressible, i.e., obeys Liouville theorem). Let $\mathbf{v}$ be the velocity vector field
$\mathbf{v}=(\dot q, \dot p, \dot s, \dot p_s)$. From (\ref{Hoover3}) follows that
$\, \text{div}\, \mathbf{v} = \partial \dot q/ \partial q+ \partial \dot p/ \partial p + \partial \dot s/ \partial s + \partial \dot p_s/ \partial p_s= p_s/\tau^2 \neq 0$, implying that the set of Eqs.  (\ref{Hoover3}) is not Hamiltonian.

In the next step Hoover makes the change of variables $v = p/s$ and $\zeta = p_s/\tau^2$ to obtain the well known
Nos\'e-Hoover equations:
\begin{equation}
\dot q = v \ ; \dot v = -q -\zeta v\ ;
 \ \dot \zeta  = [v^2 - T]/\tau ^2
\label{Hoover4}
\end{equation}
The divergence of the associated velocity field $\mathbf{v}=(\dot q, \dot v, \dot \zeta)$
is nonzero: $\, \text{div}\, \mathbf{v} = \partial \dot q/ \partial q+ \partial \dot v/ \partial v +  \partial \dot \zeta/ \partial \zeta=-\zeta\neq0$", implying that, as we state in our paper \cite{Campisi},
\emph{the Nos\'e-Hoover Equations}  (\ref{Hoover4}) \emph{are not Hamiltonian}.
This is indeed a statement that  often appears in the previous literature. Klages \cite{Klages} repeatedly
states that the Nos\'e-Hoover equations are not Hamiltonian. Kusnezov, Bulgac and  Bauer  in Ref. \cite{Kusnezov} comment about
the Nos\'e-Hoover Equations that: ``\emph{These equations of motion no longer retain a Hamiltonian structure. In the extended
phase space the variables $q,p$ and $\zeta$ are no longer canonical and the symplectic structure is lost}''.

Hoover \cite{Hoover} continues with deriving the Nos\'e-Hoover equations  (\ref{Hoover4}) from the Hamiltonian of Dettmann  \cite{Dettmann}:
\begin{equation}
H_{\text{Dettmann}}= p^2/2s+sq^2/2+sp_s^2/2\tau^2 +sT\ln s
\label{Hoover5}
\end{equation}
The corresponding canonical equations
\begin{align}
\dot q &  = p/s  \nonumber \\
\dot p & =-s q \nonumber  \\
\dot s &= sp_s/\tau^2  \nonumber  \\
\dot p_s &= p^2/2s^2 - q^2/2-p_s^2/2\tau^2-T\ln s -T \nonumber  \\
\label{Hoover6}
\end{align}
are, by definition, Hamiltonian.
Hoover  \cite{Hoover} then chooses a special value of the Hamiltonian, $H_{\text{Dettmann}}=0$, and uses this numerical value of the Hamiltonian in (\ref{Hoover6}), to obtain the set of Eqs. (\ref{Hoover3}).
We note that replacing the Hamiltonian function by the numerical value of the energy, as it is determined by the initial conditions, actually leads to equations of motion which no longer obey Liouville's theorem in that the divergence of the according vector field does not vanish. As we have seen, the set of  Eqs. (\ref{Hoover3}) is not volume preserving. Consequently the so modified equations of motion (\ref{Hoover3}) are no longer Hamiltonian even though they produce the same identical trajectories as the Hamiltonian equations of motion (\ref{Hoover6}) if  one only starts with initial conditions with the same energy value as the one which replaces the Hamiltonian in the modified equations of motion. For any initial condition with a different energy, however, the Hamiltonian equations (\ref{Hoover6}) and the modified equations (\ref{Hoover3}) will typically lead to different trajectories. The formal reason why the modified equations of motion do not satisfy Liouville's theorem lies in the fact that the divergence of a vector field at any point of phase space ``sees'' the full neighborhood of this phase space point and  also takes into account directions transversal to the energy hyper-surface.

Hoover  \cite{Hoover} also considers another Hamiltonian (setting now $\tau=1$ and $T=1$):
\begin{equation}
H_{\text{Another}} \equiv (s/2)[ \ (p/s)^2 + q^2 + p_s^2 \ ] + s\ln (s) - s
\label{Hoover10}
\end{equation}
and shows that for a specific initial condition yielding the zeroing of $H_{\text{Another}}$, 
it gives the same trajectory as the Nos\'e-Hoover equations.
Exactly as in the case discussed above, this is not sufficient to guarantee that the Nos\'e-Hoover equations are 
Hamiltonian, because just like $H_{\text{Dettmann}}$, $H_{\text{Another}}$ does produce the same trajectories
as the Nos\'e-Hoover equations only on a specific energy shell $H_{\text{Another}}=0$.

In the last two paragraphs of his comment \cite{Hoover}, Hoover emphasizes that the Nos\'e-Hoover equations, when employed to
simulate heat transport, give multifractal strange attractors in the full system+thermostat phase space. Since Hamiltonian flows
cannot possess attractors, this fact provides further evidence that the Nos\'e-Hoover equations are not Hamiltonian.

We agree with Hoover's statement, in the last paragraph of his comment \cite{Hoover}, that if our method
``is more easily matched in laboratory experiments then it is indeed a step forward''.
This is indeed the case. The Hamiltonian that we propose reads:
\begin{equation}
 H= \sum_i \frac{p_i^2}{2m_i} + V(\mathbf q) +
\frac{P^2}{2M}+T\ln\frac{|X|}{b}+ h(\mathbf q,X)
\label{eq:H}
\end{equation}
where $h(\mathbf{q},X)$ is a weak interaction term between the system ($\mathbf{q},\mathbf{p}$) and a log-oscillator ($X,P$),
and $b>0$ sets the length scale of the log-oscillator. Its major advantage
is that it can be  matched in laboratory experiments, because it presents the kinetic term in the standard form $ \sum_i {p_i^2}/{2m_i}+{P^2}/{2M}$.
In contrast, both Nos\'e's Hamiltonian (\ref{Hoover1}), Dettman's Hamiltonian (\ref{Hoover5}), as well as the Hamiltonian in
(\ref{Hoover10})  involve  non-standard kinetic terms
containing the log-oscillator position $s$, which can not readily (if at all) be matched in laboratory experiments.
Specifically, Nos\'e's Hamiltonian  (\ref{Hoover1}) contains the non-standard kinetic term $p^2/s^2$, 
whereas Dettman's Hamiltonian  (\ref{Hoover5}) and  the Hamiltonian in
(\ref{Hoover10}) contain the non-standard kinetic terms $p^2/s,sp_s^2/2\tau^2$.
As we have elucidated with our manuscript \cite{Campisi},
one way to implement the Hamiltonian (\ref{eq:H}) is to put $N$ neutral atoms and one charged ion in a box in presence
of the 2D logarithmic Coulomb potential (this can be generated by a thin long charged wire). Through short ranged collisions,
the ion, acting as thermostat, will thermalize the gas of neutral atoms to a Gibbs distribution of temperature $T$ given by the strength of the logarithmic potential (which is proportional to the charge on the ion and to the linear charge density on the wire).
Moreover, logarithmic potentials can also be created artificially by means of properly engineered laser fields for cold atoms  \cite{Schleich10PRA82} and colloidal particles \cite{Cohen05PRL94}, and occur naturally in various situations: For example it is known that the motion of stars in elliptical galaxies is governed by logarithmic potentials \cite{Stoica}, and also that a logarithmic potential determines the interaction of vortices in unsteady flows \cite{Onsager}.

On a mathematical level, the most distinctive feature of our method is that, 
at variance with the approach described by Hoover in his comment \cite{Hoover} in our approach one need not 
choose specific values of the energy in order to obtain thermostated dynamics. Our Hamiltonian (\ref{eq:H}) produces 
thermostated dynamics \emph{globally}, i.e., over the whole phase space: for this reason we
find more appropriate to qualify our thermostat as a ``truly Hamiltonian thermostat'' rather than just 
``Another Hamiltonian thermostat'' as in Hoover's comment \cite{Hoover}.

We conclude with the remark that the purpose of our work \cite{Campisi} is not to criticize or to improve over the 
Nos\'e-Hoover thermostat, whose importance and usefulness is out question.


\end{document}